\documentclass[floatfix,twocolumn,aps,showpacs,amsmath,amssymb]{revtex4}

\usepackage{graphicx}
\usepackage{dcolumn}
\usepackage{bm}

\begin{document}

\title{Alpha-–decay half--lives, alpha-–capture and alpha-–nucleus potential}

\author{V. Yu. Denisov, A. A. Khudenko}
\address{%
Institute for Nuclear Research, Prospect Nauki 47,
03680 Kiev, Ukraine }%

\date{\today}

\begin{abstract}
The $\alpha$-decay half-lives and the $\alpha$-capture cross-sections are evaluated in the framework of unified model for $\alpha$-decay and $\alpha$-capture. In the framework of this model the $\alpha$-decay and $\alpha$-capture are considered as penetration of the $\alpha$-particle through the potential barrier formed by nuclear, Coulomb and centrifugal interactions between $\alpha$-particle and nucleus. The spins and the parities of parent and daughter nuclei as well as the quadrupole and hexadecapole deformations of daughter nuclei are taken into account for evaluation of the $\alpha$-decay half-lives. The $\alpha$-decay half-lives for 344 nuclei and the $\alpha$-capture cross-sections of $^{40}$Ca, $^{44}$Ca, $^{59}$Co, $^{208}$Pb and $^{209}$Bi agree well with the experimental data. The evaluated $\alpha$-decay half-lives within the range $10^{-9}\le T_{1/2} \le 10^{38}$ s for 1246 $\alpha$-emitters are tabulated.
\end{abstract}


\maketitle

\section{Introduction}
Alpha-decay is very important process in nuclear physics \cite{gamov,cg,back,brazil,dasgupta,toi,nds,audi,nudat,gupta,belli,jaeri,karamian,she,micr,kadmenski-furman,kadmeski,stewart,delion,bm,silesteanu,strutinsky,royer,basu,blendowske,xu,brasil2,samanta,di,bg,karp,poenaru,gn,vs,mnms2,sss,sp,brown,fuji,rz}. The experimental information on $\alpha$-decay half-lives is extensive and is being continually updated (see Refs.
\cite{back,brazil,dasgupta,toi,nds,audi,nudat,gupta,belli,jaeri,karamian,she} and papers cited therein). The theory of $\alpha$-decay was formulated by Gamow \cite{gamov} and independently by Condon and Gurney \cite{cg} in 1928. Subsequently various microscopic
\cite{micr,kadmenski-furman,kadmeski,stewart,delion,bm,silesteanu}, macroscopic cluster \cite{back,strutinsky,royer,basu,blendowske,xu,brasil2,samanta,di,bg} and fission \cite{brazil,poenaru} approaches to the description of $\alpha$-decay have been proposed. The simple empirical relations for description of the $\alpha$-decay half-lives are extensively discussed too, see, for example, Refs. \cite{dasgupta,gupta,royer,poenaru,gn,vs,mnms2,sss,sp,brown,fuji,rz} and numerous references therein.

The $\alpha$-decay process involves sub-barrier penetration of $\alpha$-particles through the barrier, caused by interaction between $\alpha$-particle and nucleus. The fusion ($\alpha$-capture) reaction between $\alpha$-particle and nucleus proceeds in the opposite direction to the $\alpha$-decay reaction. However, the same $\alpha$-nucleus interaction potential is the principal factor to describe both reactions \cite{di}. Therefore it is natural to use data for both the $\alpha$-decay half-lives and the around barrier $\alpha$-capture reactions for determination of the $\alpha$-nucleus interaction potential \cite{di}. Note that  $\alpha$-decay and $\alpha$-capture has been also discussed simultaneously in Ref. \cite{bg} recently.

Now we use a combination of updated $\alpha$-decay half-lives dataset for the ground-state-to-ground-state transitions from data compilations {\it Table of Isotopes} \cite{toi}, \cite{nds}, {\it Nubase} \cite{audi}, \cite{nudat} and Ref. \cite{belli} as well as the $\alpha$-capture cross-sections of $^{40}$Ca \cite{subfus-exp-ca1,subfus-exp-ca2}, $^{44}$Ca \cite{subfus-exp-ca1}, $^{59}$Co \cite{subfus-exp-co},  $^{208}$Pb \cite{subfus-exp-pb} and $^{209}$Bi \cite{subfus-exp-pb} around barrier. We stress that the $\alpha$-decay from the ground-state of parent nucleus can proceed into both the ground-state and excited states of daughter nucleus \cite{toi,nds}. Therefore it is necessary to take into account the branching ratio of $\alpha$-decay relatively other decay modes (fission, $\beta$-decay and etc.) \cite{toi,nds,audi,nudat} as well as the branching ratio of $\alpha$-decay into the ground-state \cite{toi,nds} relatively the total $\alpha$-decay half-life during evaluation of the dataset for $\alpha$-decay half-lives for the ground-state-to-ground-state transitions from data. Carefully updated and selected $\alpha$-decay half-lives dataset contains information on reliable data for the 344 ground-state-to-ground-state $\alpha$-transitions. Note that the $\alpha$-decay half-lives data for 367 nuclei and the $\alpha$-capture cross-sections of $^{40}$Ca, $^{59}$Co and  $^{208}$Pb around barrier were used in Ref. \cite{di}. Both our datasets are wider than those considered in Ref. \cite{bg}.

By using our dataset for $\alpha$-decay half-lives and $\alpha$-capture reactions we can determine the $\alpha$-nucleus potential deeply below and around barrier with high degree of accuracy. Knowledge of the $\alpha$-nucleus interaction potential is a key for the analysis of various reactions between $\alpha$-particle and nuclei. Therefore obtained $\alpha$-nucleus potential can be used for description of various reactions in nuclear physics and astrophysics.

Many $\alpha$-emitters are deformed. Therefore $\alpha$-nucleus potential should depend on the angle $\theta$ between the direction of $\alpha$-emission and the axial-symmetry axis of the deformed nucleus. Both the $\alpha$-decay half-life and the transmission coefficient for tunneling through the barrier are strongly dependent on $\theta$ \cite{micr,kadmeski,stewart,delion,bm,strutinsky,di}, because the transmission coefficient exponentially depends on the $\alpha$-nucleus potential values. This effect is elaborately discussed in microscopic models \cite{stewart,delion,bm}. The quadrupole deformation and angle effects are considered in the cluster approach in Ref. \cite{di}, while the influence of quadrupole and hexadecapole deformations of daughter nuclei was studied in Ref. \cite{xu}. Therefore we take into account both quadrupole and hexadecapole deformations of daughter nuclei in present work.

Nuclei with stable ground state deformation are the most bounded at equilibrium shape, which is deformed. The difference between binding energies of such nuclei at deformed and spherical shapes is the deformation energy ${\cal E}_{\rm def}$ \cite{str,fh,mnms}. Note that values of ${\cal E}_{\rm def}$ are close to 5-10 MeV for well-deformed heavy nuclei \cite{str,fh,mnms}. If deformed parent and daughter nuclei are considered as spherical, then the energy balance of $\alpha$-decay should take into account variation of the deformation energy. This affects strongly on the condition of $\alpha$-emission, because $\alpha$-decay half-life is very sensitive to the variation of the energy released in $\alpha$-transition.

The interaction potential between $\alpha$-particle and nuclei consists of nuclear, Coulomb and centrifugal parts.
The nuclear and Coulomb parts are taken into account in the evaluation of the $\alpha$-decay half-lives and $\alpha$-capture cross sections in Ref. \cite{di}. However the centrifugal part of $\alpha$-nucleus potential is exactly accounted for evaluation of $\alpha$-capture cross-sections and ignored in calculation of $\alpha$-decay half-lives \cite{di}, because the spins and parities of parent and daughter nuclei as well as angular momentum of $\alpha$-transitions are neglected. Nevertheless, $\alpha$-transitions between ground states of even-odd, odd-even and odd-odd nuclei occur at non-zero values of angular momentum of the $\alpha$-particle, when the spins and/or parities of parent and daughter nuclei are different. As the result, the centrifugal potential distinctly contributes to the total $\alpha$-nucleus potential at small distances between daughter nucleus and $\alpha$-particle. The $\alpha$-decay half-life depends exponentially on the action, which is very sensitive to the $\alpha$-nucleus potential. Therefore accurate consideration of the $\alpha$-transitions should take into account the spins and parities of parent and daughter nuclei and angular momentum of the emitted $\alpha$-particle \cite{kadmenski-furman,bm}.

The experimental values and theoretical estimates of the ground-state spins and parities are known for many nuclei \cite{audi,nudat}. Moreover the number of nuclei with known values of ground-state spin and parity is permanently extended. Therefore we revalue the $\alpha$-nucleus interaction potential using available updated data for $\alpha$-decay half-lives, the spins and parities of the ground-states of parent and daughter nuclei and $\alpha$-capture reaction cross-sections. Due to this our approach becomes more accurate.

Our unified model for $\alpha$-decay and $\alpha$-capture ({\it UMADAC}) is shortly discussed in Sec. 2. The selection of adjustable parameters and discussion of the results are given in Sec. 3. The Sec. 4 is dedicated to conclusions.

\section{Unified model for $\alpha$-decay and $\alpha$-capture}

The $\alpha$-decay half-life $T_{1/2}$ is calculated as \cite{di}
\begin{eqnarray}
T_{1/2} = \hbar \ln(2)/\Gamma,
\end{eqnarray}
where
\begin{eqnarray}
\Gamma = \frac{1}{4\pi} \int \gamma(\theta,\phi) d\Omega
\end{eqnarray}
is the total width of decay, $\gamma(\theta,\phi)$ is the partial width of $\alpha$-emission in direction $\theta$ and $\phi$ and $\Omega$ is the space angle.

The width of $\alpha$-emission in direction $\theta$ for axial-symmetric nuclei is given as the following:
\begin{eqnarray}
\gamma(\theta) = \hbar \; 10^\nu \; t(Q_\alpha,\theta,\ell),
\end{eqnarray}
where $10^{\nu}$ is the $\alpha$-particle frequency assaults the barrier, which takes into account the $\alpha$-particle preformation, $t(Q_\alpha,\theta,\ell)$ is the transmission coefficient, which shows the probability of penetration through the barrier, and $Q_\alpha$ is the released energy at $\alpha$-decay.

The transmission coefficient can be obtained in the semiclassical WKB approximation
\begin{eqnarray}
\lefteqn{t(Q_\alpha,\theta,\ell) = 1/\{ 1} \\ &+& \left.  \exp\left[\frac{2}{\hbar}
\int_{a(\theta)}^{b(\theta)} dr \sqrt{2\mu
\left(v(r,\theta,\ell,Q_\alpha)-Q_\alpha\right)} \right] \right\}, \nonumber
\end{eqnarray}
where $a(\theta)$ and $b(\theta)$ are the inner and outer turning points determined from the equations
$v(r,\theta,\ell,Q_\alpha)|_{r=a(\theta),b(\theta)}=Q_\alpha$, and $\mu$ is the reduced mass. The $\alpha$-nucleus potential $v(r,\theta,\ell,Q_\alpha)$ consists of Coulomb $v_\mathrm{C}(r,\theta)$, nuclear $v_\mathrm{N}(r,\theta,Q_\alpha)$ and centrifugal $v_\ell(r)$ parts, i.e.
\begin{eqnarray}
v(r,\theta,\ell,Q_\alpha)=v_\mathrm{C}(r,\theta) + v_\mathrm{N}(r,\theta,Q_\alpha) + v_\mathrm{\ell}(r),
\end{eqnarray}
where
\begin{equation}
v_\mathrm{C}(r,\theta) = \frac{2 Z e^2}{r} \left[1 + \frac{3R^2}{5r^2}\beta_{2} Y_{20}(\theta) \right.
+\left. \frac{3R^4}{9r^4}\beta_{4} Y_{40}(\theta) \right]
\end{equation}
for $r \ge r_\mathrm{c}(\theta)$,
\begin{eqnarray}
v_\mathrm{C}(r,\theta)&\approx&\frac{2 Z e^2}{r_\mathrm{c}(\theta)}\left[\frac{3}{2}-\frac{r^2}{2r_\mathrm{c}(\theta)^2} \right. \hspace{3.5cm}\nonumber\\
& +&\left. \frac{3R^2}{5r_\mathrm{c}(\theta)^2} \beta_{2} Y_{20}(\theta) \left(2-\frac{r^3}{r_\mathrm{c}(\theta)^3} \right) \right. \nonumber\\
& +& \left. \frac{3R^4}{9r_\mathrm{c}(\theta)^4}\beta_{4} Y_{40}(\theta)\left(\frac{7}{2}-\frac{5r^2}{2r_\mathrm{c}(\theta)^2}\right) \right]
\end{eqnarray}
for $r \le r_\mathrm{c}(\theta)$,
\begin{equation}
\hspace{-2.2cm} v_\mathrm{N}(r,\theta,Q_\alpha) = \frac{V(Q_\alpha)}{1+\exp[(r-r_\mathrm{m}(\theta))/d]},
\end{equation}
\begin{equation}
\hspace{-5cm} v_\mathrm{\ell}(r) =\frac{\hbar^2 \ell (\ell+1)}{2\mu r^2}.
\end{equation}
Here $Z$, $R$, $\beta_{2}$ and $\beta_{4}$ are, respectively, the number of protons, the radius, the quadrupole and hexadecapole deformation parameters of the nucleus interacting with the $\alpha$-particle; $e$ is the charge of proton, $Y_{20}(\theta)$ and $Y_{40}(\theta)$ are harmonic functions; $V(Q_\alpha)$ and $r_\mathrm{m}(\theta)$ are, correspondingly, the strength and effective radius of the nuclear part of $\alpha$-nucleus potential. The inner turning point $a(\theta)$ is close to both $r_\mathrm{m}(\theta)$ and $r_\mathrm{c}(\theta)$. Presentation of Coulomb field in the form (6) at distances $r \lesssim r_\mathrm{c}(\theta)$ ensures the continuity of the Coulomb field and its derivative at $r=r_\mathrm{c}(\theta)$. We choose  $r_\mathrm{c}(\theta)=r_\mathrm{m}(\theta)$ to reduce the number of parameters. Note that Eq. (6) describes the Coulomb potential between spherical and deformed nuclei at distances, when interacting nuclei are separated \cite{dp}. By substituting $\beta_2=\beta_4=0$ we reduce Eq. (7) to the well-known form of the Coulomb potential into uniformly-charged sphere.

The $\alpha$-particle emission from nuclei obeys the spin-parity selection rule. Let $j_\mathrm{p}, \pi_\mathrm{p}$ and $j_\mathrm{d}, \pi_\mathrm{d}$ are the spin and parity values of the parent and daughter nuclei respectively. The $\alpha$-particle has zero value of spin and positive parity, therefore the minimal value of angular momentum $\ell_{\rm min}$ at the $\alpha$-transition between states with $j_\mathrm{p}, \pi_\mathrm{p}$ and $j_\mathrm{d}, \pi_\mathrm{d}$ is
\begin{equation}
\ell_{\rm min} = \left\{
\begin{array}{llll}
\Delta_j & {\rm for} \; {\rm even} & \Delta_j & {\rm and} \;\; \pi_\mathrm{p} = \pi_\mathrm{d}, \\
\Delta_j+1 & {\rm for} \; {\rm odd} & \Delta_j & {\rm and} \;\; \pi_\mathrm{p} = \pi_\mathrm{d}, \\
\Delta_j & {\rm for} \; {\rm odd} & \Delta_j & {\rm and} \;\; \pi_\mathrm{p} \neq \pi_\mathrm{d} , \\
\Delta_j+1 & {\rm for} \; {\rm even} & \Delta_j & {\rm and} \;\; \pi_\mathrm{p} \neq \pi_\mathrm{d},
\end{array} \right.
\end{equation}
where $\Delta_j=|j_\mathrm{p}-j_\mathrm{d}|$.

Note that the value of $\alpha$-particle angular momentum $\ell$ can be larger than $\ell_{\rm min}$. This is related to the intrinsic structure of the single-particle levels around proton and neutron Fermi levels in parent and daughter nuclei and the way of $\alpha$-particle formation in parent nuclei. There are many cases of $\alpha$-transition between ground states with non-zero value of angular momentum. We suppose that the angular momentum of $\alpha$-transition between ground states $\ell$ equals to $\ell_{\rm min}$ for the sake of simplicity. So, the centrifugal part of the $\alpha$-nucleus potential (see Eq. (9)) is determined according to the spin-parity selection rule for $\alpha$-transition. The centrifugal contribution to the potential is very important for $\alpha$-emission from even-odd, odd-even and odd-odd nuclei. We consider that $\ell_{\rm min} = 0$ for all even-even nuclei.

The $\alpha$-capture cross section of axial-symmetric nucleus at around-barrier collision energy $E$ in the center-of-mass system is equal to \cite{di}
\begin{eqnarray}
\sigma(E)=\frac{\pi \hbar^2}{2\mu E} \int_0^{\pi/2} \sum_\ell
(2\ell+1) t(E,\theta,\ell) \sin(\theta) d\theta.
\end{eqnarray}
Here the integration over angle $\theta$ is done for the same reason as in Eq. (2). The transmission coefficient $t(E,\theta,\ell)$ can be evaluated using the semiclassical WKB approximation (see Eq. (4)) in the case of collision between $\alpha$-particle and stiff magic or near-magic spherical nuclei at collision energies $E$ below and slightly above barrier. The $\alpha$-nucleus potential is given by Eqs. (5)--(9). The transmission coefficient is approximated by an expression for a parabolic barrier at collision energies higher than or equal to the barrier energy. This approximation for the transmission coefficient is very common in the case of sub-barrier fusion reaction between heavy ions \cite{di,subfus-rev,denisov-tr}.

\section{Discussion and results}

\subsection{Input data}

We chose data for $T_{1/2}$ for 344 $\alpha$-decay transitions between the ground states of parent and daughter nuclei with exact values of the half-lives, the $\alpha$-decay branching ratio relatively the other decay modes (fission, $\beta$-decay and etc.) and the branching ratio of ground-state-to-ground-state $\alpha$-decay transitions relatively $\alpha$-decay transitions from the ground-state of parent nucleus to excited states of daughter nucleus from Tables in Refs. \cite{toi,nds,audi,nudat} and add data from recent paper \cite{belli}. The $\alpha$-decay half-lives marked in Refs. \cite{audi,nudat} as poorly estimated or with limit for half-life  have not been included into our database. So we have selected only well-defined  ground-state-to-ground-state $\alpha$-transitions. (However our selection rule is not so strict as the one in Ref. \cite{dasgupta}.) As the result, 136 even-even, 84 even-odd, 76 odd-even and 48 odd-odd $\alpha$-emitters are included in the database. The selected dataset of $\alpha$-emitters has very large mass $106 \leq A \leq 261$ and charge $52 \leq Z \leq 107$ ranges. Due to selection procedure the number of $\alpha$-decay half-lives for even-even nuclei considered now is slightly smaller than the one in Refs. \cite{royer,xu,di}, but much larger than the one in Refs. \cite{dasgupta}. Note that 77 $\alpha$-emitters within narrow ranges $146 \leq A \leq 255$ and $62 \leq Z \leq 100$ are accounted for parameter search due to very strict selection rule applied in Ref. \cite{dasgupta}.

The released energy of $\alpha$-particle emitted from nucleus in $\alpha$-decay is calculated using recent evaluation of atomic mass data \cite{audi}. The effect of atomic electrons on the energy of $\alpha$-particle should be also taken into account. Therefore the released energy of $\alpha$-particle $Q_\alpha$ emitted from nucleus in $\alpha$-decay is \cite{brasil2,huang}
\begin{eqnarray}
Q_\alpha=\Delta {\cal M}_\mathrm{p} -(\Delta {\cal M}_\mathrm{d}+\Delta M_\mathrm{\alpha})+ k(Z_\mathrm{p}^\epsilon-Z_\mathrm{d}^\epsilon) ,
\end{eqnarray}
where $\Delta {\cal M}_\mathrm{p}$, $\Delta {\cal M}_\mathrm{d}$ and $\Delta M_\mathrm{\alpha}$ are, correspondingly, the mass-excess of parent, daughter and $\alpha$ nuclei. The last term in Eq. (12) describes the effect of atomic electrons, $kZ^\epsilon$ represents the total binding energy of $Z$ electrons in the atom, {\em k}=8.7 eV and $\epsilon$=2.517 for nuclei with $Z\ge60$ and {\em k}=13.6 eV and $\epsilon$=2.408 for nuclei with $Z<60$ \cite{brasil2,huang}.

The experimental data on deformation parameters $\beta_{2}$ and $\beta_{4}$ are taken from the {\it RIPL-2} database \cite{ripl}. When no experimental data exist for a nuclide in the {\it RIPL-2} compilation, values of the deformation parameters are taken from the macroscopic-microscopic model \cite{mnms}.

The ground-state-to-ground-state $\alpha$-transitions of
even-even nuclei take place at $\ell=0$. The value of $\ell$ for the ground-state-to-ground-state transitions in even-odd, odd-even and odd-odd nuclei are determined by the spin-parity selection rule, see Eq. (10). The values of spin and parity for nuclei are taken from \cite{audi}. When no data exist for a nuclide in Ref. \cite{audi}, we use corresponding values from \cite{nudat}. Unfortunately, there are cases, when the values of spin and parity are absent in both Refs. \cite{audi} and \cite{nudat}. For such nuclei we assign to the spin and parity values $0^+$ in our calculations and leave a empty space in the Table II.

The data for $\alpha$-capture cross sections of $^{40}$Ca, $^{44}$Ca, $^{59}$Co, $^{208}$Pb and $^{209}$Bi were taken from Refs.
\cite{subfus-exp-ca1,subfus-exp-ca2,subfus-exp-co,subfus-exp-pb}. We consider $\alpha$-capture cross sections using the same approach as in Ref. \cite{di}. (Note that we take into account data points for $\alpha$-capture cross sections of $^{40}$Ca, $^{44}$Ca for below and near barrier energies, because at high collision energies other processes can become important, as the result, the one-dimensional model for $\alpha$-capture is not proper.) Therefore we shortly discuss $\alpha$-capture reactions below.

Note that unified analysis of the experimental data for both $\alpha$-decay and $\alpha$-capture gives unique possibility to evaluate the mass $A$, charge $Z$ and energy $Q_\alpha$ dependencies of the $\alpha$-nucleus potential in the very wide ranges $40 \leq A \leq 293$, $50 \leq Z \leq 118$ and  1.915 MeV $\leq Q_\alpha \leq$ 25 MeV. The obtained mass, charge and energy dependencies of the $\alpha$-nucleus potential can be applied in wider ranges and for various purposes too.

\subsection{Parameter search}

We want to describe both the half-lives for ground-state-to-ground-state $\alpha$-transitions in 344 nuclei and $\alpha$-capture cross-sections of $^{40}$Ca (twosets), $^{44}$Ca, $^{59}$Co, $^{208}$Pb and $^{209}$Bi using the {\it UMADAC} presented in Sec. 2. By solving this task we parametrize $V(Q_\alpha)$, $r_\mathrm{m}(\theta)$, $d$ and $\nu$ in Eqs. (3)--(9) and determine these parameters by searching the minimum of function
\begin{eqnarray}
\hspace{-0.6cm}F&=& \left( 5 D_\mathrm{{e-e}}+D_\mathrm{{e-o}}+D_\mathrm{{o-e}}+D_\mathrm{{o-o}} \right)\nonumber\\
&+& 20 \left(3 D_\sigma^{\rm ^{208}Pb} +3 D_\sigma^{\rm ^{209}Bi}  +
D_\sigma^{^{40}Ca,1} + D_\sigma^{^{40}Ca,2} \right. \\
& +& \left.  D_\sigma^{^{44}Ca} + D_\sigma^{\rm ^{59}Co}\right).\nonumber
\end{eqnarray}
Here
\begin{eqnarray}
D_\mathrm{e-e}&=&\sum_\mathrm{{e-e}}
\left[\log_{10}(T_{1/2}^{\rm theor})-\log_{10}(T_{1/2}^{\rm exp})\right]^2 \hspace{1.5cm}\\
&=& \sum_\mathrm{{e-e}}\left[{\cal T}^{\rm theor}-{\cal T}^{\rm exp} \right]^2
\nonumber
\end{eqnarray}
is the difference between decimal logarithm of theoretical $T_{1/2}^{\rm theor}$ and experimental $T_{1/2}^{\rm exp}$ values of $\alpha$-decay half-lives for set of even-even nuclei, ${\cal T}^\mathrm{theor} = \log_{10}(T_{1/2}^\mathrm{theor})$, ${\cal T}^\mathrm{exp} = \log_{10}(T_{1/2}^\mathrm{exp})$, $D_\mathrm{{e-o}}$, $D_\mathrm{{o-e}}$, $D_\mathrm{{o-o}}$ are the differences similar to (14) for even-odd, odd-even and odd-odd datasets respectively,
\begin{eqnarray}
D_\sigma=\sum_{k} \left[ \log_{10}(\sigma^{\rm theor}(E_k)) - \log_{10}(\sigma^{\rm exp}(E_k)) \right]^2.
\end{eqnarray}
Here $\sigma^{\rm theor}(E_k)$ and $\sigma^{\rm exp}(E_k)$ are, correspondingly, the theoretical and experimental values of $\alpha$-capture cross section of corresponding nucleus at energy $E_k$.

By inserting various coefficients in Eq. (13) we take into account that
\begin{itemize}
\item[--] $\alpha$-decay half-lives data are known better than data for $\alpha$-capture reactions, as a rule;
\item[--] description of $\alpha$-decay half-lives in even-even nuclei is the most accurate in the framework of our model, because there is no the angular momentum uncertainty for such $\alpha$-transitions;
\item[--] the value of $D_\mathrm{e-e}$ is several times smaller then values of $D_\mathrm{e-o}$, $D_\mathrm{o-e}$ or $D_\mathrm{o-o}$, however both the data and our description of $\alpha$-decay half-lives in even-even nuclei are the most accurate, therefore we choose factor 5 in the first line of Eq. (13) for the sake of reinforcing the role of even-even nuclei during parameter search (note that $5D_\mathrm{e-e} \approx 2/3(D_\mathrm{e-o} + D_\mathrm{o-e} +D_\mathrm{o-o})$);
\item[--] the cross sections for different $\alpha$-capture reactions \cite{subfus-exp-ca1,subfus-exp-ca2,subfus-exp-co,subfus-exp-pb} are known with different accuracy. (Moreover, two experimental datasets available for reaction $\alpha+^{40}$Ca \cite{subfus-exp-ca1,subfus-exp-ca2} are in poor agreement with each other.) Taking into account that $D_\sigma^{\rm ^{208}Pb} ( {\rm or} \; D_\sigma^{\rm ^{209}Bi}) <<
D_\sigma^{^{40}Ca,1} ( {\rm or}  \; D_\sigma^{^{40}Ca,2} , {\rm or} \; D_\sigma^{^{44}Ca}, {\rm or} \; D_\sigma^{\rm ^{59}Co})$ and the cross sections for $\alpha$-capture on $^{208}$Pb and $^{209}$Bi are known with highest accuracy we introduce factor 3 in the second line of Eq. (13), which enhances the influence of $\alpha$-capture data on $^{208}$Pb and $^{209}$Bi at parameter search.
\end{itemize}
It is reasonable that the contribution of $\alpha$-capture reactions into function $F$ was close to 10\%, therefore we multiply by 20 the contribution of $\alpha$-capture reactions (see Eq. (13)).

As the result of minimization for various forms of parameters $V(Q_\alpha)$, $r_\mathrm{m}(\theta)$ and $d$ in Eqs. (6)--(9) we find the minimum of the function $F$ (13) at
\begin{eqnarray}
V(Q_\alpha)&=&v_1 + \frac{v_2 Z}{A^{1/3}} + v_3 I
+ \frac{v_4 Q_\alpha}{A^{1/3}}+\frac{v_5 Y_{20}(\theta) \beta_2}{A^{1/6}},
\\
r_\mathrm{m}(\theta) &=&r_1 + R (1+\beta_{2} Y_{20}(\theta) + \beta_{4} Y_{40}(\theta)),
\\
R&=&r_2 A^{1/3} (1 + r_3 /A+ r_4 I),\\
d&=&d_1+d_2A^{-1/3}, \\
\nu&=&19+S+\nu_0 Z^{1/2}A^{1/6}+\nu_1 ((-1)^{\ell}-1) \\
&+&\nu_2\frac{Z}{\sqrt{Q_\alpha}}+\nu_3 I+\nu_4 \beta_{2}+\nu_5 \beta_{4}+\nu_6 \frac{\ell(\ell+1)}{A^{1/6}} \nonumber
\end{eqnarray}
where $A$ and $Z$ are the number of nucleons and protons in nucleus, which is interacting with $\alpha$-particle, $I=(A-2Z)/A=(N-Z)/A$, $S=4.1382$, $S=3.57016$, $S=3.8246$ and $S=3.6625$ for even-even, even-odd, odd-even and odd-odd nuclei correspondingly. Note that 22 parameters are contained in Eqs. (16)-(20). The parameters values are given in Table I.

\begin{table}[h]
\caption{The parameters of the $\alpha$-nucleus potential and the assault frequency.}
%\renewcommand{\tabcolsep}{1.2 pc} % enlarge column spacing
%\renewcommand{\arraystretch}{1.2} % enlarge line spacing
%\vspace{0.2cm}
\begin{tabular}{p{3.7cm} p{3.7cm}}
\hline
\hline
$v_1$ (MeV) & -40.1031 \\
$v_2$ (MeV) & $-10.1511 \times 10^{-2}$\\
$v_3$ (MeV) &  -9.1928\\
$v_4$  &  $10.8545 \times 10^{-5}$\\
$v_5$ (MeV) & $6.0703 \times 10^{-2}$ \\
$r_1$ (fm) & 1.1683 \\
$r_2$ (fm) &  1.2915\\
$r_3$  & 1.4088 \\
$r_4$  & -0.0994 \\
$d_1$ (fm) & 0.6870 \\
$d_2$ (fm) & -0.3664 \\
$\nu_0$ (s) & -0.1348 \\
$\nu_{1}$ & 0.9132\\
$\nu_2$ (MeV$^{-1/2}$) & $-4.1029 \times 10^{-2}$ \\
$\nu_3$  &  0.6564\\
$\nu_4$  & -1.6442 \\
$\nu_5$  &  -1.2112\\
$\nu_6$  &  $6.8513 \times 10^{-2}$\\
\hline
\hline\\
\end{tabular}
\end{table}

The strength of the nuclear part of the interaction potential depends on the Coulomb parameter $Z/A^{1/3}$, the proton-neutron symmetry $I$ and the reaction energy $Q_\alpha$.
The angular and deformation dependences of the interaction strength (see the last term in Eq. (16)) reflect the fact that the strength of nuclear part of potential between spherical and deformed nuclei is smaller for tip orientation of deformed nucleus and larger for side orientation  \cite{dn,dp}. We also introduce the quadrupole and hexadecapole deformation dependences of the factor $\nu$(see Eq. (20)). The deformation dependence of the factor $\nu$ shows that the formation of the $\alpha$-particle on the surface of deformed parent nucleus is hindered in comparison with spherical parent nucleus, and the assault frequency is reduced in deformed nuclei in comparison with the spherical ones due to enlargement of the mean surface radius as a result of surface deformation. Various values of parameter $S$ for even-even, even-odd, odd-even and odd-odd nuclei are related to hindrance of $\alpha$-particle formation on the surface of even-odd, odd-even and, especially, odd-odd parent nuclei. Moreover, $\alpha$-particle preformation should be influenced by parity or spin of the $\alpha$-transition, see factors $\nu_1$ and $\nu_6$ correspondingly.

The results of $\alpha$-decay half-lives and $\alpha$-capture cross-sections evaluated in the framework of our {\it UMADAC} are presented below. We start our discussion with detailed consideration of the $\alpha$-decay half-lives.

\subsection{$\alpha$-decay half-lives}

The evaluated $\alpha$-decay half-lives agree well with 344 experimental data points, see  Fig. 1 and Tables 1 and 2. The experimental values of half-lives are scattered over an extremely wide range from $\sim 10^{-8}$ s to $\sim 10^{27}$ s. The $\alpha$-decay half-lives are very nicely described in the case of even-even parent nuclei. We see in Fig. 1 that the difference between theoretical and experimental values of $\log_{10}T_{1/2}$ are smaller than 0.4 for most of cases of even-even nuclei and smaller than 0.8 for most of cases of even-odd, odd-even and odd-odd nuclei.

We present the $\alpha$-decay half-lives between the ground states of parent and daughter nuclei obtained in the framework of our {\it UMADAC} in Table II. All possible $\alpha$-emitters with evaluated $\alpha$-decay half-lives within the range $10^{-9}$ s $ \le T_{1/2} \le 10^{38}$ s are included into Table II. As the result, there are 1246 $\alpha$-emitters in Table II, among them 344 and 902 $\alpha$-emitters with known and unknown values of the $\alpha$-decay half-life respectively. Note that the $T_{1/2} = (1.9 \pm 0.2) \times 10^{19}$ yr = $(6.0 \pm 0.6) \times 10^{26}$ s is the longest $T_{1/2}$ value for $\alpha$-decays observed so far \cite{audi,belli}. Therefore our upper limit for $T_{1/2}\le 10^{38}$ s gives an adequate margin for planning experiments in foreseeable future.

The root-mean-square (rms) error of decimal logarithm of $\alpha$-decay half-lives is determined as
\begin{eqnarray}
\hspace{-0.6cm}\delta &=& \sqrt{\frac{1}{N-1} \sum_\mathrm{{k=1}}^\mathrm{N} \left[\log_{10}(T_{1/2}^{\rm theor})-\log_{10}(T_{1/2}^{\rm exp})\right]^2} .
\end{eqnarray}
We use this expression for evaluation of the total $\delta_\mathrm{tot}$ and partial (even-even $\delta_\mathrm{e-e}$, even-odd $\delta_\mathrm{e-o}$, odd-even $\delta_\mathrm{o-e}$ and odd-odd $\delta_\mathrm{o-o}$) rms errors in the framework of our and other models by using our dataset for $T_{1/2}^{\rm exp}$. The values of rms errors $\delta_\mathrm{tot}$, $\delta_\mathrm{e-e}$, $\delta_\mathrm{e-o}$, $\delta_\mathrm{o-e}$ and $\delta_\mathrm{o-o}$ obtained in our model are presented in Tables III-IV. We see in Tables III-IV that the values of these errors are small.

\subsection{$\alpha$-capture cross-sections}

The $\alpha$-capture cross-sections of $^{40}$Ca, $^{44}$Ca, $^{59}$Co, $^{208}$Pb and $^{209}$Bi evaluated using Eqs. (4)-(11), (16)-(20) are compared with experimental data \cite{subfus-exp-ca1,subfus-exp-ca2,subfus-exp-co,subfus-exp-pb} in Fig. 2. We see that the data for $\alpha$-capture of $^{208}$Pb and $^{209}$Bi are precisely described in the framework of the {\it UMADAC}. The cross section for $\alpha$-capture of $^{40}$Ca is well reproduced at low energies and slightly overestimated at higher energies. In contrast to this the cross sections for $\alpha$-capture of $^{44}$Ca and $^{59}$Co are well reproduced at high energies and slightly overestimated at very low energies.

In the framework of {\it UMADAC} a one-dimensional model for evaluation of the fusion cross-section between an $\alpha$-particle and a spherical nucleus is used. It is well-known that the coupled-channel effects are very important for the nucleus-nucleus fusion reaction around the barrier \cite{di,subfus-rev,denisov-tr,CCFULL,ccdef}. Thus, we also made result of the coupled-channel calculation of the fusion cross-section for reaction $\alpha$+$^{208}$Pb by using {\it CCFULL} code \cite{CCFULL}, and presented the results on Fig. 2. The effects of nonlinear coupling of the low-energy surface vibrational states in all orders are taken into account in this code. The {\it CCFULL} calculation uses the same $\alpha$-nucleus potential as in the case of one-dimensional calculation. The values of excitation energies and surface deformations are taken from \cite{ripl}. As we can see in Fig. 2, the agreement between our one-dimensional and coupled-channel calculations is very good. The good agreement between {\it CCFULL} and one-dimensional calculations can be attributed to the high stiffness of double-magic nuclei participating in this reaction. Note that due to this reason we select for our consideration $\alpha$-capture reactions of $^{40}$Ca, $^{44}$Ca, $^{59}$Co, $^{208}$Pb and $^{209}$Bi. All these nuclei are very stiff.

\subsection{Comparison with other approaches}

The $\alpha$-decay is considered in recent Refs. \cite{she,royer,xu,brasil2,sss,sp,fuji}. Our results and those from Ref. \cite{xu} are obtained by
different cluster model approaches to the $\alpha$-decay, while results from Refs. \cite{royer,sss} are evaluated with the help of various empirical relations. The empirical relations used in Refs. \cite{dasgupta,gupta,royer,brasil2,poenaru,gn,vs,mnms2,sss,sp,brown,fuji,rz} and in numerous references cited in these Refs. couple $\log_{10}(T_{1/2})$ to $\alpha$-particle energy $Q_\alpha$,  mass $A$ and charge $Z$ of parent nuclei by simple functional expressions, i.e  $\log_{10}(T_{1/2})=f(Q_\alpha,A,Z)$, where $f(Q_\alpha,A,Z)$ is function.

As a rule, empirical relationships are derived by using a pure Coulomb picture of $\alpha$-decay, which neglects
\begin{itemize} \item[--] the nuclear force between $\alpha$-particle and daughter nucleus, \item[--] the deformation of daughter nucleus and \item[--] the spin and parity values of $\alpha$-transitions.\end{itemize} The empirical relationships are based on the fitting parameters and special analytical expressions, which are similar to the Viola--Seaborg \cite{vs} relationship. The empirical relationships are often used to estimate  $\log_{10}(T_{1/2})$ due to their simplicity and acceptable accuracy. The empirical relationship from Ref. \cite{sss} was derived especially for description of $\log_{10}(T_{1/2})$ in heavy and superheavy nuclei. In Ref. \cite{royer} four empirical relationships for even-even, even-odd, odd-even and odd-odd $\alpha$-decaying nuclei are established.

We compare values of the rms errors of decimal logarithm of $\alpha$-decay half-lives $\delta_\mathrm{tot}$, $\delta_\mathrm{e-e}$, $\delta_\mathrm{e-o}$, $\delta_\mathrm{o-e}$ and $\delta_\mathrm{o-o}$ obtained in the framework of our {\it UMADAC} and other models \cite{dasgupta,royer,brasil2,mnms2} in Table III. All values of the rms errors are evaluated for our dataset for $T_{1/2}^{\rm exp}$. The lowest values of the rms errors of decimal logarithm of $\alpha$-decay half-lives for any set of nuclei are obtained in our approach. The spectacular reduction of the rms errors $\delta_\mathrm{e-o}$, $\delta_\mathrm{o-e}$ and $\delta_\mathrm{o-o}$ in our model is obtained due to careful consideration of the spin-parity selection rules. It should be noted here, that the values of rms errors for some relationships, which are given in original papers related to corresponding relationships, can deviate from values presented in Table III, because different datasets for experimental $\alpha$-decay half-lives are used in the original papers. The datasets difference may be caused by three reasons.
\begin{itemize}
\item[--] Various nuclei are included into the datasets of different papers.
\item[--] Different values for $\alpha$-decay half-lives and/or $Q_\alpha$ are included into the datasets from various experiments. Note that experimental data are permanently improved and extended. \item[--] The data for $\alpha$-decay half-lives for the ground-state-to-ground-state transitions are only included into our dataset in contrast to some other datasets.
\end{itemize}

The empirical relationships gives reasonable accuracy for the $\alpha$-transitions in even-even nuclei, because the angular momentum of the ground-state-to-ground-state $\alpha$-transition equals zero. However the empirical relationships are too rough for even-odd, odd-even and odd-odd $\alpha$-emitters, because the angular momentum of $\alpha$-transition in such nuclei is often non-zero.

Some empirical relationships are established for very heavy $\alpha$-emitters. Therefore we compare values of the rms errors $\delta_\mathrm{tot}$, $\delta_\mathrm{e-e}$, $\delta_\mathrm{e-o}$, $\delta_\mathrm{o-e}$ and $\delta_\mathrm{o-o}$ obtained in our {\it UMADAC} and other models \cite{dasgupta,gupta,royer,brasil2,mnms2,fuji,rz} for $A \geq 208$ and $Z \geq 82$ in Table IV. In this case we select 144 $\alpha$-emitters, among them 59  even-even, 33 even-odd, 34 odd-even and 18 odd-odd $\alpha$-emitters. The lowest values of the rms errors $\delta_\mathrm{tot}$, $\delta_\mathrm{e-o}$, $\delta_\mathrm{o-e}$ and $\delta_\mathrm{o-o}$ are obtained in our model, however the value of the rms error $\delta_\mathrm{e-e}$ evaluated in our model is not the lowest one. The values of rms errors for very heavy $\alpha$-emitters are larger than corresponding ones for the total dataset. This is probably related to the fact that the $\alpha$-decay energy, spins and parities of parent and daughter nuclei are worse known for very heavy $\alpha$-emitters. Due to this more accurate experimental information on the values of the mass excess, spin, parity and deformations of the ground-state of nuclei can help to improve both the accuracy and predicted reliability of our model. The other reason is related to the fact that some relationships are established by fitting data for very heavy $\alpha$-emitters, and therefore these relations are better for such range of $\alpha$-emitters.

\section{Conclusions}

We determined the $\alpha$-nucleus potential by
using the data for $\alpha$-decay half-lives for 344 $\alpha$-emitters and around-barrier $\alpha$-capture cross sections of $^{40}$Ca, $^{44}$Ca, $^{59}$Co, $^{208}$Pb and $^{209}$Bi.

In the framework of the {\it UMADAC} we take into account deformation and spin-parity effects in evaluation of the $\alpha$-decay half-lives.

The data for $\alpha$-decay half-lives of 344 spherical and deformed nuclei and for $\alpha$-capture cross sections of $^{40}$Ca, $^{44}$Ca, $^{59}$Co, $^{208}$Pb and $^{209}$Bi are well described in the framework of the {\it UMADAC}.

We predict $\alpha$-decay half-lives for the ground-state-to-ground-state transitions in 902 nuclei.

By taking into account the spins and parities of parent and daughter nuclei we obtain spectacular improvement in description of the $\alpha$-decay half-lives in even-odd, odd-even and odd-odd nuclei.

\section*{Acknowledgments}

The authors thanks V. I. Tretyak for useful remarks. The authors are indebted to the Referee for helpful suggestions.

\clearpage
%\onecolumn

\section*{\small Explanation of tables}
\addcontentsline{toc}{section}{Explanation of Tables}

\subsection*{Table II \qquad $\alpha$-decay half-lives for  the ground-state-to-ground-state $\alpha$-transitions.}
The decimal logarithm of $\alpha$-decay half-lives for 1246 ground-state-to-ground-state $\alpha$-transitions with evaluated half-lives in the range $10^{-9}$ s $\le T_{1/2} \le 10^{38}$ s are presented in the Table. Available experimental data for $\alpha$-decay half-lives for 344 nuclei from Refs. \cite{audi,gupta,belli} are given too. The following notations were used:

\begin{table}[h]
%\caption {The result table.}
%\renewcommand{\tabcolsep}{1.5cm} % enlarge column spacing
\renewcommand{\arraystretch}{1.2} % enlarge line spacing
\begin{tabular}{ll}
$A$ & The mass number of the parent nucleus\\
$Z$ & The proton number of the parent nucleus\\
${\cal T}^\mathrm{exp}$ &The decimal logarithm of the experimental $\alpha$-decay \\
                        &half-life ${\cal T}^\mathrm{exp}=\log_{10}(T_{1/2}^\mathrm{exp})$. \\
                        &The value of $T_{1/2}^\mathrm{exp}$ is given in s.\\
${\cal T}^\mathrm{theor}$ &The decimal  logarithm of the evaluated $\alpha$-decay \\
                          &half-life ${\cal T}^\mathrm{theor}=\log_{10}(T_{1/2}^\mathrm{theor})$. \\
                          & The value of $T_{1/2}^\mathrm{theor}$ is given in s. \\
$\beta_2$ &Quadrupole deformation of the daughter nucleus\\
$\beta_4$ & Hexadecapole deformation of the daughter nucleus\\
$J_p^\pi$ &Spin and parity of the parent nucleus\\
$J_d^\pi$ &Spin and parity of the daughter nucleus\\
$l_\mathrm{min}$ &The minimal orbital momentum of the emitted \\
                 &$\alpha$-particle evaluated according to the selection rule (10). \\
\end{tabular}
\end{table}

1246 nuclei are listed in the Table II, i.e. all possible $\alpha$-emitters with evaluated half-lives in the range $10^{-9}$ s $\le T_{1/2} \le 10^{38}$ s are included in Table.

If the experimental values of the $\alpha$-decay half-lives are not known, than we put a dash.

The energies of ground-state-to-ground-state $\alpha$-transitions can be easy evaluated using the evaluated atomic mass data \cite{audi} and Eq. (11). Due to this the energies of $\alpha$-transitions are not given here.

The spin and parity values are presented according to notations of Ref. \cite{audi}. The spin and parity values  without and with parentheses are based upon strong and weak assignment arguments \cite{audi}, respectively. The mark $\#$ indicates spin and/or parity values estimated from systematic trends in neighboring nuclides with the same $N$ and $Z$.

By default values of half-lives, spins and parities are extracted from Ref. \cite{audi}.  However there are specials cases, which marked in Table by letters: \\ $a$ - the spin value is extracted from Ref. \cite{nudat} while the parity value is picked up from Ref. \cite{audi}; \\ $b$ - the spin value is taken from Ref. \cite{audi} and the parity value is adopted from Ref. \cite{nudat}; \\ $c$ - both the spin and parity values are taken from Ref. \cite{nudat}. \\ We hold an empty place for unknown spins and parities. As the result, we substitute $l_\mathrm{min}=0$ for such cases.

The values of the quadrupole and hexadecapole deformation of the daughter nuclei are taken from Refs. \cite{mnms,ripl}, see text for details. \\

\vspace{-0.7cm}
\subsection*{Table III \qquad RMS errors of the decimal logarithm of $\alpha$-decay half-lives for full set of $\alpha$-emitters.}

The rms error of the decimal logarithm of $\alpha$-decay half-lives is evaluated according to Eq. (21). The full set of $\alpha$-emitters with known half-life values contains 344 nuclei, among them 136 even-even, 84 even-odd, 76 odd-even and 48 odd-odd nuclei. The experimental half-life values were taken from Refs. \cite{audi,gupta,belli}. The first line is result of our {\it UMADAC}, while other lines are evaluated by using various relationships. The last column contains the References for corresponding relationships. \\

\vspace{-0.7cm}
\subsection*{Table IV \qquad  RMS errors of the decimal logarithm of $\alpha$-decay half-lives for nuclei heavier than lead $^{208}_{82}$Pb. }

The set of $\alpha$-emitters heavier than lead with known half-life values contains 144 nuclei, among them 59 even-even, 33 even-odd, 34 odd-even and 18 odd-odd nuclei. The experimental values were picked up from Refs. \cite{audi,gupta}. The notations in Table IV are similar to the ones of Table III. \\

\vspace{-0.5cm}
\section*{The explanation of Figures}
\addcontentsline{toc}{section}{Explanation of Figures}

\vspace{-0.2cm}
\subsection*{Figure 1 \qquad Comparison between
experimental and theoretical values of $\log_{10}(T_{1/2})$  for $\alpha$-decays.}

Left panels: The experimental (circles)
\cite{audi,nudat,gupta,belli} and theoretical (crosses) values of
$\log_{10}(T_{1/2})$ for $\alpha$-decays in even-even (e-e),
even-odd (e-o), odd-even (o-e) and odd-odd (o-o) parent nuclei.\\
Right panels: Dots represent the difference between the
experimental and theoretical values of $\log_{10}(T_{1/2})$ for $\alpha$-decays
in even-even (e-e), even-odd (e-o), odd-even (o-e)
and odd-odd (o-o) parent nuclei.

The $T_{1/2}^\alpha$ values are given in s.

\vspace{-0.2cm}
\subsection*{Figure 2 \qquad The experimental and theoretical values of $\alpha$-capture cross-section of $^{40}$Ca, $^{44}$Ca, $^{59}$Co, $^{208}$Pb and $^{209}$Bi.}

Squares are data for reaction $\alpha$+$^{208}$Pb from Ref. \cite{subfus-exp-pb}, circles are data for reaction $\alpha$+$^{59}$Co from Ref. \cite{subfus-exp-co}, up- and down-pointing triangles are data for reaction $\alpha$+$^{40}$Ca from Ref. \cite{subfus-exp-ca2} and
\cite{subfus-exp-ca1}, respectively, rhombuses are data for reaction $\alpha$+$^{44}$Ca from Ref. \cite{subfus-exp-ca1}, right-pointing triangles are data for reaction $\alpha$+$^{209}$Bi from Ref. \cite{subfus-exp-pb}. Lines are results of calculations obtained in the framework of {\it UMADAC} and stars are result of calculations using {\it CCFULL} code \cite{CCFULL}.

\newpage
%\twocolumn

%\section*{Tables}

\vspace{-1 cm}
\begin{table}
\vspace{0.3 cm}
\caption {$\alpha$-decay half-lives for the ground-state-to-ground-state $\alpha$-transitions}
\vspace{0.4 cm}
%\label{table:1}
\renewcommand{\tabcolsep}{0.2 pc} % enlarge column spacing
\renewcommand{\arraystretch}{1.2} % enlarge line spacing
\small
\begin{tabular}{ccccccccc}
\hline
\hline
$A_p$ & $Z_p$ & ${\cal T}^{\mathrm{theor}}$ & ${\cal T}^{\mathrm{exp}}$ & $\beta_2$ & $\beta_4$ & $j_\mathrm{p}^{\pi}$ & $j_\mathrm{d}^{\pi}$& $l_\mathrm{min}$ \\
\hline
103	&	51	&	6.00	&	-	&	0.035	&	-0.015	&	$2.5^+\#$	&	$4.5^+\#$	&	 2	\\
104	&	51	&	7.03	&	-	&	0.053	&	0.001	&	$$	&	$(6)^+$	&	0	\\
105	&	51	&	9.15	&	-	&	0.053	&	-0.007	&	$(2.5^+)$	&	$4.5^+\#$	&	 2	\\
105	&	52	&	-5.75	&	-	&	0.027	&	0.024	&	$2.5^+\#$	&	$2.5^+\#$	&	 0	\\
106	&	51	&	14.69	&	-	&	0.054	&	-0.014	&	$(4^+)$	&	$(6^+)$	&	2	 \\
106	&	52	&	-4.61	&	-4.15	&	0.009	&	-0.015	&	$0^+$	&	$0^+$	&	0	 \\
107	&	51	&	22.39	&	-	&	0.053	&	-0.007	&	$2.5^+\#$	&	$4.5^+\#$	&	 2	\\
\hline
\hline
\end{tabular}
\end{table}

\begin{table}
%\caption {Something}
%\label{table:1}
\renewcommand{\tabcolsep}{0.2 pc} % enlarge column spacing
\renewcommand{\arraystretch}{1.2} % enlarge line spacing
\small
\begin{tabular}{ccccccccc}
\hline
\hline
$A_p$ & $Z_p$ & ${\cal T}^{\mathrm{theor}}$ & ${\cal T}^{\mathrm{exp}}$ & $\beta_2$ & $\beta_4$ & $j_\mathrm{p}^{\pi}$ & $j_\mathrm{d}^{\pi}$& $l_\mathrm{min}$ \\
\hline
118	&	56	&	13.70	&	-	&	0.221	&	0.052	&	$0^+$	&	$0^+$	&	0	 \\
118	&	57	&	12.42	&	-	&	0.252	&	0.084	&	$$	&	$(1^+)$	&	0	\\
119	&	55	&	24.83	&	-	&	0.197	&	0.039	&	$4.5^+$	&	$2.5^+\#$	&	2	 \\
119	&	56	&	25.17	&	-	&	0.243	&	0.064	&	$(2.5^+)$	&	$(2.5^+)$	&	 0	\\
119	&	57	&	16.49	&	-	&	0.252	&	0.075	&	$5.5^-\#$	&	$4.5^+\#$	&	 1	\\
119	&	58	&	12.00	&	-	&	0.271	&	0.095	&	$2.5^+\#$	&	$2.5^+\#$	&	 0	\\
120	&	56	&	22.67	&	-	&	0.249	&	0.046	&	$0^+$	&	$0^+$	&	0	 \\
\hline
\hline
\end{tabular}
\end{table}

\begin{table}
%\caption {Something}
%\label{table:1}
\renewcommand{\tabcolsep}{0.2 pc} % enlarge column spacing
\renewcommand{\arraystretch}{1.2} % enlarge line spacing
\small
\begin{tabular}{ccccccccc}
\hline
\hline
$A_p$ & $Z_p$ & ${\cal T}^{\mathrm{theor}}$ & ${\cal T}^{\mathrm{exp}}$ & $\beta_2$ & $\beta_4$ & $j_\mathrm{p}^{\pi}$ & $j_\mathrm{d}^{\pi}$& $l_\mathrm{min}$ \\
\hline
133	&	62	&	14.31	&	-	&	0.322	&	-0.003	&	$(2.5^+)$	&	$2.5^+\#$	&	 0	\\
133	&	63	&	11.14	&	-	&	0.33	&	0.026	&	$5.5^-\#$	&	$2.5^+\#$	&	 3	\\
134	&	60	&	36.99	&	-	&	0.258	&	-0.018	&	$0^+$	&	$0^+$	&	0	 \\
134	&	61	&	23.17	&	-	&	0.283	&	-0.012	&	$(5^+)$	&	$(6^{+\#})$	&	2	 \\
134	&	62	&	13.79	&	-	&	0.37	&	0.002	&	$0^+$	&	$0^+$	&	0	 \\
134	&	63	&	9.56	&	-	&	0.331	&	0.007	&	$$	&	$(4^+)$	&	0	\\
134	&	64	&	6.72	&	-	&	0.331	&	0	&	$0^+$	&	$0^+$	&	0	\\
\hline
\hline
\end{tabular}
\end{table}

\begin{table}
%\caption {Something}
%\label{table:1}
\renewcommand{\tabcolsep}{0.2 pc} % enlarge column spacing
\renewcommand{\arraystretch}{1.2} % enlarge line spacing
\small
\begin{tabular}{ccccccccc}
\hline
\hline
$A_p$ & $Z_p$ & ${\cal T}^{\mathrm{theor}}$ & ${\cal T}^{\mathrm{exp}}$ & $\beta_2$ & $\beta_4$ & $j_\mathrm{p}^{\pi}$ & $j_\mathrm{d}^{\pi}$& $l_\mathrm{min}$ \\
\hline
144	&	65	&	23.69	&	-	&	0.171	&	-0.039	&	$1^+$	&	$1^+$	&	0	 \\
144	&	66	&	15.97	&	-	&	0.21	&	-0.043	&	$0^+$	&	$0^+$	&	0	 \\
144	&	67	&	16.40	&	-	&	0.228	&	-0.04	&	$$	&	$5$	&	0	\\
144	&	68	&	10.18	&	-	&	0.267	&	-0.05	&	$0^+$	&	$0^+$	&	0	 \\
145	&	60	&	31.25	&	-	&	-0.035	&	0.009	&	$3.5^-$	&	$3.5^-$	&	0	 \\
145	&	61	&	17.92	&	17.30	&	0	&	0	&	$2.5^+$	&	$2.5^+$	&	0	\\
145	&	65	&	35.05	&	-	&	-0.156	&	-0.022	&	$(1.5^+)$	&	$2.5^+$	&	2	 \\
\hline
\hline
\end{tabular}
\end{table}

\begin{table}
%\caption {Something}
%\label{table:1}
\renewcommand{\tabcolsep}{0.2 pc} % enlarge column spacing
\renewcommand{\arraystretch}{1.2} % enlarge line spacing
\small
\begin{tabular}{ccccccccc}
\hline
\hline
$A_p$ & $Z_p$ & ${\cal T}^{\mathrm{theor}}$ & ${\cal T}^{\mathrm{exp}}$ & $\beta_2$ & $\beta_4$ & $j_\mathrm{p}^{\pi}$ & $j_\mathrm{d}^{\pi}$& $l_\mathrm{min}$ \\
\hline
150	&	70	&	10.72	&	-	&	-0.173	&	-0.035	&	$0^+$	&	$0^+$	&	0	 \\
150	&	71	&	9.09	&	-	&	-0.199	&	-0.038	&	$(5^-)$	&	$(6^-)$	&	2	 \\
151	&	63	&	25.55	&	26.20	&	0.161	&	0.059	&	$2.5^+$	&	$3.5^+$	&	2	 \\
151	&	64	&	16.28	&	15.03	&	0.143	&	0.056	&	$3.5^-$	&	$3.5^-$	&	0	 \\
151	&	65	&	8.54	&	8.82	&	0	&	0	&	$0.5^{(+)}$	&	$2.5^+$	&	2	 \\
151	&	66	&	4.55	&	4.28	&	0.116	&	0.009	&	$3.5^{(-)}$	&	$3.5^-$	&	 0	\\
151	&	67	&	4.29	&	-	&	-0.008	&	0	&	$5.5^{(-)}$	&	$0.5^+\#$	&	5	 \\
\hline
\hline
\end{tabular}
\end{table}

\begin{table}
%\caption {Something}
%\label{table:1}
\renewcommand{\tabcolsep}{0.2 pc} % enlarge column spacing
\renewcommand{\arraystretch}{1.2} % enlarge line spacing
\small
\begin{tabular}{ccccccccc}
\hline
\hline
$A_p$ & $Z_p$ & ${\cal T}^{\mathrm{theor}}$ & ${\cal T}^{\mathrm{exp}}$ & $\beta_2$ & $\beta_4$ & $j_\mathrm{p}^{\pi}$ & $j_\mathrm{d}^{\pi}$& $l_\mathrm{min}$ \\
\hline
156	&	73	&	3.00	&	-	&	-0.079	&	-0.013	&	$(2^-)$	&	$(5^-)$	&	4	 \\
157	&	67	&	29.17	&	-	&	0.216	&	0.043	&	$3.5^-$	&	$2.5^+$	&	1	 \\
157	&	68	&	12.52	&	-	&	0.17	&	0.044	&	$1.5^-$	&	$3.5^{(-)}$	&	2	 \\
157	&	69	&	10.46	&	-	&	0.152	&	0.041	&	$0.5^+$	&	$5.5^-$	&	5	 \\
157	&	70	&	4.27	&	3.89	&	0.107	&	0.037	&	$3.5^-$	&	$3.5^{(-)}$	&	 0	\\
157	&	71	&	4.40	&	-	&	-0.018	&	0	&	$(0.5^+)$	&	$(5.5^-)$	&	5	 \\
157	&	72	&	-0.88	&	-0.91	&	-0.052	&	0.009	&	$3.5^-$	&	$3.5^{-c}$	&	 0	\\
\hline
\hline
\end{tabular}
\end{table}

\begin{table}
%\caption {Something}
%\label{table:1}
\renewcommand{\tabcolsep}{0.2 pc} % enlarge column spacing
\renewcommand{\arraystretch}{1.2} % enlarge line spacing
\small
\begin{tabular}{ccccccccc}
\hline
\hline
$A_p$ & $Z_p$ & ${\cal T}^{\mathrm{theor}}$ & ${\cal T}^{\mathrm{exp}}$ & $\beta_2$ & $\beta_4$ & $j_\mathrm{p}^{\pi}$ & $j_\mathrm{d}^{\pi}$& $l_\mathrm{min}$ \\
\hline
163	&	74	&	1.90	&	0.83	&	0.134	&	0.031	&	$1.5^-\#$	&	$3.5^{-c}$	 &	2	\\
163	&	75	&	-0.09	&	-0.22	&	0.107	&	0.012	&	$(0.5^+)$	&	$(0.5^+)$	 &	0	\\
163	&	76	&	-1.87	&	-	&	0.08	&	0.002	&	$3.5^-\#$	&	$3.5^-\#$	&	 0	\\
164	&	69	&	29.33	&	-	&	0.272	&	0.044	&	$1^+$	&	$5^+$	&	4	 \\
164	&	70	&	20.65	&	-	&	0.304	&	0.04	&	$0^+$	&	$0^+$	&	0	 \\
164	&	71	&	15.01	&	-	&	0.225	&	0.028	&	$1^{(-)}$	&	$1^-$	&	0	 \\
164	&	72	&	9.20	&	-	&	0.23	&	0.016	&	$0^+$	&	$0^+$	&	0	 \\
\hline
\hline
\end{tabular}
\end{table}

\begin{table}
%\caption {Something}
%\label{table:1}
\renewcommand{\tabcolsep}{0.2 pc} % enlarge column spacing
\renewcommand{\arraystretch}{1.2} % enlarge line spacing
\small
\begin{tabular}{ccccccccc}
\hline
\hline
$A_p$ & $Z_p$ & ${\cal T}^{\mathrm{theor}}$ & ${\cal T}^{\mathrm{exp}}$ & $\beta_2$ & $\beta_4$ & $j_\mathrm{p}^{\pi}$ & $j_\mathrm{d}^{\pi}$& $l_\mathrm{min}$ \\
\hline
169	&	73	&	11.79	&	-	&	0.245	&	-0.002	&	$(2.5^+)$	&	$0.5^+$	&	2	 \\
169	&	74	&	8.43	&	-	&	0.217	&	0.009	&	$(2.5^-)$	&	$(2.5^-)$	&	 0	\\
169	&	75	&	4.36	&	-	&	0.199	&	-0.01	&	$4.5^-\#$	&	$2.5^-\#$	&	 2	\\
169	&	76	&	1.83	&	1.59	&	0.17	&	0.002	&	$1.5^-\#$	&	$1.5^-\#$	 &	0	\\
169	&	77	&	0.22	&	-0.11	&	0.153	&	0	&	$0.5^+\#$	&	$0.5^+\#$	&	 0	\\
169	&	78	&	-1.52	&	-	&	0.125	&	0.006	&	$1.5^-\#$	&	$(3.5^-)$	&	 2	\\
169	&	79	&	-3.33	&	-	&	0.099	&	-0.012	&	$0.5^+\#$	&	$0.5^+\#$	&	 0	\\
\hline
\hline
\end{tabular}
\end{table}

\begin{table}
%\caption {Something}
%\label{table:1}
\renewcommand{\tabcolsep}{0.2 pc} % enlarge column spacing
\renewcommand{\arraystretch}{1.2} % enlarge line spacing
\small
\begin{tabular}{ccccccccc}
\hline
\hline
$A_p$ & $Z_p$ & ${\cal T}^{\mathrm{theor}}$ & ${\cal T}^{\mathrm{exp}}$ & $\beta_2$ & $\beta_4$ & $j_\mathrm{p}^{\pi}$ & $j_\mathrm{d}^{\pi}$& $l_\mathrm{min}$ \\
\hline
174	&	75	&	10.62	&	-	&	0.254	&	-0.001	&	$$	&	$(3^{+\#})$	&	0	 \\
174	&	76	&	5.27	&	5.34	&	0.24	&	-0.006	&	$0^+$	&	$0^+$	&	0	 \\
174	&	77	&	2.64	&	-	&	0.199	&	-0.019	&	$(3^+)$	&	$(5^+)$	&	2	 \\
174	&	78	&	0.12	&	0.03	&	0.171	&	-0.014	&	$0^+$	&	$0^+$	&	0	 \\
174	&	79	&	-0.82	&	-0.81	&	0.153	&	-0.007	&	$$	&	$$	&	0	\\
174	&	80	&	-2.73	&	-2.70	&	0.107	&	-0.004	&	$0^+$	&	$0^+$	&	0	 \\
175	&	72	&	26.24	&	-	&	0.295	&	-0.024	&	$2.5^-$	&	$0.5^-$	&	2	 \\
\hline
\hline
\end{tabular}
\end{table}

\begin{table}
%\caption {Something}
%\label{table:1}
\renewcommand{\tabcolsep}{0.2 pc} % enlarge column spacing
\renewcommand{\arraystretch}{1.2} % enlarge line spacing
\small
\begin{tabular}{ccccccccc}
\hline
\hline
$A_p$ & $Z_p$ & ${\cal T}^{\mathrm{theor}}$ & ${\cal T}^{\mathrm{exp}}$ & $\beta_2$ & $\beta_4$ & $j_\mathrm{p}^{\pi}$ & $j_\mathrm{d}^{\pi}$& $l_\mathrm{min}$ \\
\hline
179	&	79	&	1.43	&	-	&	0.171	&	-0.022	&	$2.5^-\#$	&	$(2.5^-)$	&	 0	\\
179	&	80	&	1.07	&	-	&	0.18	&	0.004	&	$2.5^-\#$	&	$3.5^{-c}$	&	 2	\\
179	&	81	&	-0.55	&	-0.57	&	-0.122	&	-0.01	&	$(0.5^+)$	&	$0.5^+\#$	 &	0	\\
179	&	82	&	-2.71	&	-	&	-0.105	&	-0.019	&	$2.5^-\#$	&	$2.5^-\#$	&	 0	\\
180	&	73	&	36.75	&	-	&	0.278	&	-0.071	&	$1^+$	&	$7^-$	&	7	 \\
180	&	74	&	25.60	&	25.75	&	0.295	&	-0.062	&	$0^+$	&	$0^+$	&	0	 \\
180	&	75	&	19.14	&	-	&	0.277	&	-0.045	&	$(1)^-$	&	$(1)^-$	&	0	 \\
\hline
\hline
\end{tabular}
\end{table}

\begin{table}
%\caption {Something}
%\label{table:1}
\renewcommand{\tabcolsep}{0.2 pc} % enlarge column spacing
\renewcommand{\arraystretch}{1.2} % enlarge line spacing
\small
\begin{tabular}{ccccccccc}
\hline
\hline
$A_p$ & $Z_p$ & ${\cal T}^{\mathrm{theor}}$ & ${\cal T}^{\mathrm{exp}}$ & $\beta_2$ & $\beta_4$ & $j_\mathrm{p}^{\pi}$ & $j_\mathrm{d}^{\pi}$& $l_\mathrm{min}$ \\
\hline
185	&	79	&	5.44	&	4.99	&	0.238	&	-0.038	&	$2.5^-$	&	$(2.5)^-$	&	 0	\\
185	&	80	&	3.18	&	2.93	&	0.265	&	-0.015	&	$0.5^-$	&	$0.5^-$	&	0	 \\
185	&	81	&	5.52	&	-	&	-0.148	&	-0.007	&	$0.5^+\#$	&	$(1.5^-)$	&	 1	\\
185	&	82	&	0.25	&	2.32	&	-0.122	&	-0.018	&	$1.5^-$	&	$0.5^{-a}$	&	 2	\\
185	&	83	&	-4.33	&	-	&	-0.053	&	-0.007	&	$0.5^{+c}$	&	$0.5^+\#$	&	 0	\\
186	&	75	&	34.78	&	-	&	0.259	&	-0.101	&	$1^-$	&	$3^-$	&	2	 \\
186	&	76	&	22.79	&	22.80	&	0.251	&	-0.084	&	$0^+$	&	$0^+$	&	0	 \\
\hline
\hline
\end{tabular}
\end{table}

\begin{table}
%\caption {Something}
%\label{table:1}
\renewcommand{\tabcolsep}{0.2 pc} % enlarge column spacing
\renewcommand{\arraystretch}{1.2} % enlarge line spacing
\small
\begin{tabular}{ccccccccc}
\hline
\hline
$A_p$ & $Z_p$ & ${\cal T}^{\mathrm{theor}}$ & ${\cal T}^{\mathrm{exp}}$ & $\beta_2$ & $\beta_4$ & $j_\mathrm{p}^{\pi}$ & $j_\mathrm{d}^{\pi}$& $l_\mathrm{min}$ \\
\hline
191	&	81	&	11.90	&	-	&	-0.156	&	-0.014	&	$(0.5^+)$	&	$0.5^+$	&	0	 \\
191	&	82	&	5.64	&	-	&	-0.139	&	-0.024	&	$(1.5^-)$	&	$1.5^-$	&	0	 \\
191	&	83	&	2.39	&	2.85	&	-0.053	&	-0.007	&	$(4.5^-)$	&	$(0.5^+)$	 &	5	\\
191	&	84	&	-1.41	&	-	&	0	&	-0.015	&	$1.5^-\#$	&	$(1.5^-)$	&	0	 \\
192	&	78	&	30.62	&	-	&	0.186	&	-0.086	&	$0^+$	&	$0^+$	&	0	 \\
192	&	79	&	21.98	&	-	&	0.173	&	-0.071	&	$1^-$	&	$1^-$	&	0	 \\
192	&	80	&	19.33	&	-	&	0.186	&	-0.013	&	$0^+$	&	$0^+$	&	0	 \\
\hline
\hline
\end{tabular}
\end{table}

\begin{table}
%\caption {Something}
%\label{table:1}
\renewcommand{\tabcolsep}{0.2 pc} % enlarge column spacing
\renewcommand{\arraystretch}{1.2} % enlarge line spacing
\small
\begin{tabular}{ccccccccc}
\hline
\hline
$A_p$ & $Z_p$ & ${\cal T}^{\mathrm{theor}}$ & ${\cal T}^{\mathrm{exp}}$ & $\beta_2$ & $\beta_4$ & $j_\mathrm{p}^{\pi}$ & $j_\mathrm{d}^{\pi}$& $l_\mathrm{min}$ \\
\hline
199	&	84	&	4.13	&	3.44	&	-0.026	&	-0.015	&	$(1.5^-)$	&	$1.5\#^-$	 &	0	\\
199	&	85	&	0.97	&	0.90	&	-0.052	&	0.009	&	$(4.5^-)$	&	$(4.5^-)$	 &	0	\\
199	&	86	&	0.59	&	-	&	0.071	&	0.002	&	$1.5^-\#$	&	$1.5^-\#$	&	 0	\\
199	&	87	&	-2.32	&	-	&	-0.215	&	0.009	&	$0.5^+\#$	&	$0.5^{+c}$	&	 0	\\
200	&	82	&	23.42	&	-	&	0.116	&	-0.032	&	$0^+$	&	$0^+$	&	0	 \\
200	&	83	&	12.92	&	-	&	-0.071	&	-0.014	&	$7^+$	&	$2^-$	&	5	 \\
200	&	84	&	3.71	&	3.66	&	0	&	-0.008	&	$0^+$	&	$0^+$	&	0	 \\
\hline
\hline
\end{tabular}
\end{table}

\begin{table}
%\caption {Something}
%\label{table:1}
\renewcommand{\tabcolsep}{0.2 pc} % enlarge column spacing
\renewcommand{\arraystretch}{1.2} % enlarge line spacing
\small
\begin{tabular}{ccccccccc}
\hline
\hline
$A_p$ & $Z_p$ & ${\cal T}^{\mathrm{theor}}$ & ${\cal T}^{\mathrm{exp}}$ & $\beta_2$ & $\beta_4$ & $j_\mathrm{p}^{\pi}$ & $j_\mathrm{d}^{\pi}$& $l_\mathrm{min}$ \\
\hline
207	&	86	&	4.09	&	3.43	&	0	&	-0.015	&	$2.5^-$	&	$2.5^-$	&	0	 \\
207	&	87	&	1.29	&	1.19	&	0.045	&	-0.008	&	$4.5^-$	&	$4.5^-$	&	0	 \\
207	&	88	&	0.50	&	0.42	&	-0.104	&	0.004	&	$(1.5^-)$	&	$(1.5)^-$	 &	0	\\
207	&	89	&	-1.61	&	-	&	-0.19	&	-0.018	&	$4.5^-\#$	&	$4.5^-\#$	&	 0	\\
208	&	83	&	28.55	&	-	&	-0.044	&	-0.023	&	$(5)^+$	&	$2^-$	&	3	 \\
208	&	84	&	7.36	&	-	&	0.041	&	-0.008	&	$0^+$	&	$0^+$	&	0	 \\
208	&	85	&	5.80	&	6.04	&	-0.044	&	0.009	&	$6^+$	&	$6^+$	&	0	 \\
\hline
\hline
\end{tabular}
\end{table}

\begin{table}
%\caption {Something}
%\label{table:1}
\renewcommand{\tabcolsep}{0.2 pc} % enlarge column spacing
\renewcommand{\arraystretch}{1.2} % enlarge line spacing
\small
\begin{tabular}{ccccccccc}
\hline
\hline
$A_p$ & $Z_p$ & ${\cal T}^{\mathrm{theor}}$ & ${\cal T}^{\mathrm{exp}}$ & $\beta_2$ & $\beta_4$ & $j_\mathrm{p}^{\pi}$ & $j_\mathrm{d}^{\pi}$& $l_\mathrm{min}$ \\
\hline
213	&	86	&	-1.17	&	-1.71	&	-0.008	&	0	&	$(4.5^+)$	&	$0.5^-$	&	5	 \\
213	&	87	&	1.62	&	1.54	&	-0.035	&	-0.015	&	$4.5^-$	&	$4.5^-$	&	0	 \\
213	&	88	&	2.70	&	2.66	&	-0.044	&	-0.015	&	$0.5^-$	&	$2.5^-$	&	2	 \\
213	&	89	&	0.42	&	-0.14	&	-0.044	&	-0.015	&	$4.5^-\#$	&	$4.5^-$	&	 0	\\
213	&	90	&	-0.64	&	-0.85	&	-0.087	&	-0.005	&	$2.5^-\#$	&	$2.5^-$	&	 0	\\
213	&	91	&	-2.34	&	-	&	-0.122	&	-0.002	&	$4.5^-\#$	&	$(4.5^-)$	&	 0	\\
214	&	82	&	32.86	&	-	&	-0.026	&	-0.008	&	$0^+$	&	$0^+$	&	0	 \\
\hline
\hline
\end{tabular}
\end{table}

\begin{table}
%\caption {Something}
%\label{table:1}
\renewcommand{\tabcolsep}{0.2 pc} % enlarge column spacing
\renewcommand{\arraystretch}{1.2} % enlarge line spacing
\small
\begin{tabular}{ccccccccc}
\hline
\hline
$A_p$ & $Z_p$ & ${\cal T}^{\mathrm{theor}}$ & ${\cal T}^{\mathrm{exp}}$ & $\beta_2$ & $\beta_4$ & $j_\mathrm{p}^{\pi}$ & $j_\mathrm{d}^{\pi}$& $l_\mathrm{min}$ \\
\hline
219	&	84	&	4.26	&	-	&	0.011	&	0.01	&	$3.5^+\#$	&	$2.5^+\#$	&	 2	\\
219	&	85	&	2.65	&	-	&	-0.026	&	0.016	&	$2.5^-\#$	&	$(4.5^-)$	&	 2	\\
219	&	86	&	0.63	&	0.70	&	0.02	&	0.018	&	$2.5^+$	&	$4.5^+$	&	2	 \\
219	&	87	&	-1.06	&	-1.69	&	0.018	&	0.016	&	$4.5^-$	&	$4.5^-$	&	0	 \\
219	&	88	&	-2.52	&	-1.48	&	0.019	&	0.016	&	$(3.5)^+$	&	$4.5^+$	&	 2	\\
219	&	89	&	-4.31	&	-4.93	&	0.008	&	0.008	&	$4.5^-$	&	$4.5^-$	&	0	 \\
219	&	90	&	-5.27	&	-5.98	&	-0.018	&	0.008	&	$4.5^+\#$	&	$4.5^+\#$	 &	0	\\
\hline
\hline
\end{tabular}
\end{table}

\begin{table}
%\caption {Something}
%\label{table:1}
\renewcommand{\tabcolsep}{0.2 pc} % enlarge column spacing
\renewcommand{\arraystretch}{1.2} % enlarge line spacing
\small
\begin{tabular}{ccccccccc}
\hline
\hline
$A_p$ & $Z_p$ & ${\cal T}^{\mathrm{theor}}$ & ${\cal T}^{\mathrm{exp}}$ & $\beta_2$ & $\beta_4$ & $j_\mathrm{p}^{\pi}$ & $j_\mathrm{d}^{\pi}$& $l_\mathrm{min}$ \\
\hline
225	&	92	&	-0.34	&	-1.14	&	0.102	&	0.072	&	$2.5^+\#$	&	$(3.5^+)$	 &	2	\\
225	&	93	&	-2.52	&	-	&	-0.026	&	0.009	&	$4.5^-\#$	&	$4.5^-$	&	0	 \\
226	&	87	&	16.69	&	-	&	0.137	&	0.099	&	$1^-$	&	$$	&	0	\\
226	&	88	&	11.28	&	10.73	&	0.142	&	0.1	&	$0^+$	&	$0^+$	&	0	 \\
226	&	89	&	8.32	&	9.25	&	0.138	&	0.109	&	$(1^{-\#})$	&	$2^-$	&	 2	\\
226	&	90	&	3.58	&	3.39	&	0.192	&	0.092	&	$0^+$	&	$0^+$	&	0	 \\
226	&	91	&	2.53	&	2.45	&	0.129	&	0.099	&	$$	&	$1^-$	&	0	 \\
\hline
\hline
\end{tabular}
\end{table}

\begin{table}
%\caption {Something}
%\label{table:1}
\renewcommand{\tabcolsep}{0.2 pc} % enlarge column spacing
\renewcommand{\arraystretch}{1.2} % enlarge line spacing
\small
\begin{tabular}{ccccccccc}
\hline
\hline
$A_p$ & $Z_p$ & ${\cal T}^{\mathrm{theor}}$ & ${\cal T}^{\mathrm{exp}}$ & $\beta_2$ & $\beta_4$ & $j_\mathrm{p}^{\pi}$ & $j_\mathrm{d}^{\pi}$& $l_\mathrm{min}$ \\
\hline
232	&	95	&	2.71	&	-	&	0.191	&	0.123	&	$$	&	$$	&	0	\\
233	&	89	&	26.66	&	-	&	0.197	&	0.106	&	$(0.5^+)$	&	$0.5^+\#$	&	 0	\\
233	&	90	&	22.16	&	-	&	0.189	&	0.114	&	$0.5^+$	&	$2.5^{(+)}$	&	2	 \\
233	&	91	&	19.05	&	-	&	0.189	&	0.114	&	$1.5^-$	&	$(1.5^+)$	&	1	 \\
233	&	92	&	13.57	&	12.77	&	0.19	&	0.114	&	$2.5^+$	&	$2.5^+$	&	0	 \\
233	&	93	&	9.34	&	-	&	0.19	&	0.114	&	$(2.5^+)$	&	$(2.5^+)$	&	 0	\\
233	&	94	&	6.05	&	-	&	0.191	&	0.123	&	$2.5^+\#$	&	$(1.5^+)$	&	 2	\\
\hline
\hline
\end{tabular}
\end{table}

\clearpage
\begin{table}
%\caption {Something}
%\label{table:1}
\renewcommand{\tabcolsep}{0.2 pc} % enlarge column spacing
\renewcommand{\arraystretch}{1.2} % enlarge line spacing
\small
\begin{tabular}{ccccccccc}
\hline
\hline
$A_p$ & $Z_p$ & ${\cal T}^{\mathrm{theor}}$ & ${\cal T}^{\mathrm{exp}}$ & $\beta_2$ & $\beta_4$ & $j_\mathrm{p}^{\pi}$ & $j_\mathrm{d}^{\pi}$& $l_\mathrm{min}$ \\
\hline
238	&	98	&	0.59	&	-	&	0.216	&	0.102	&	$0^+$	&	$0^+$	&	0	 \\
239	&	91	&	30.77	&	-	&	0.215	&	0.085	&	$(1.5^{-\#})$	&	$0.5^+\#$	 &	1	\\
239	&	92	&	22.01	&	-	&	0.215	&	0.093	&	$2.5^+$	&	$0.5^+\#$	&	2	 \\
239	&	93	&	18.19	&	-	&	0.215	&	0.102	&	$2.5^+$	&	$(1.5^-)$	&	1	 \\
239	&	94	&	14.21	&	-	&	0.215	&	0.11	&	$0.5^+$	&	$3.5^-$	&	3	 \\
239	&	95	&	10.41	&	11.11	&	0.215	&	0.11	&	$(2.5)^-$	&	$2.5^+$	&	 1	\\
239	&	96	&	7.92	&	-	&	0.215	&	0.11	&	$(3.5^-)$	&	$(2.5^+)$	&	 1	\\
\hline
\hline
\end{tabular}
\end{table}

\begin{table}
%\caption {Something}
%\label{table:1}
\renewcommand{\tabcolsep}{0.2 pc} % enlarge column spacing
\renewcommand{\arraystretch}{1.2} % enlarge line spacing
\small
\begin{tabular}{ccccccccc}
\hline
\hline
$A_p$ & $Z_p$ & ${\cal T}^{\mathrm{theor}}$ & ${\cal T}^{\mathrm{exp}}$ & $\beta_2$ & $\beta_4$ & $j_\mathrm{p}^{\pi}$ & $j_\mathrm{d}^{\pi}$& $l_\mathrm{min}$ \\
\hline
245	&	95	&	12.83	&	-	&	0.224	&	0.079	&	$(2.5)^+$	&	$(2.5^+)$	&	 0	\\
245	&	96	&	11.19	&	-	&	0.224	&	0.079	&	$3.5^+$	&	$2.5^+$	&	2	 \\
245	&	97	&	6.94	&	9.37	&	0.223	&	0.087	&	$1.5^-$	&	$2.5^-$	&	2	 \\
245	&	98	&	4.12	&	3.94	&	0.223	&	0.087	&	$0.5^{(+)b}$	&	 $0.5^+$	&	0	 \\
245	&	99	&	3.73	&	3.52	&	0.224	&	0.079	&	$(1.5^-)$	&	$(3.5^+)$	 &	3	\\
245	&	100	&	2.66	&	-	&	0.224	&	0.079	&	$0.5^+\#$	&	$3.5^-\#$	&	 3	\\
245	&	101	&	-0.92	&	-	&	0.224	&	0.079	&	$0.5^-\#$	&	$(1.5^-)$	&	 2	\\
\hline
\hline
\end{tabular}
\end{table}

\begin{table}
%\caption {Something}
%\label{table:1}
\renewcommand{\tabcolsep}{0.2 pc} % enlarge column spacing
\renewcommand{\arraystretch}{1.2} % enlarge line spacing
\small
\begin{tabular}{ccccccccc}
\hline
\hline
$A_p$ & $Z_p$ & ${\cal T}^{\mathrm{theor}}$ & ${\cal T}^{\mathrm{exp}}$ & $\beta_2$ & $\beta_4$ & $j_\mathrm{p}^{\pi}$ & $j_\mathrm{d}^{\pi}$& $l_\mathrm{min}$ \\
\hline
252	&	97	&	12.14	&	-	&	0.235	&	0.039	&	$$	&	$$	&	0	\\
252	&	98	&	8.08	&	8.01	&	0.297	&	0.04	&	$0^+$	&	$0^+$	&	0	 \\
252	&	99	&	8.48	&	7.83	&	0.235	&	0.04	&	$(5^-)$	&	$6^+\#$	&	1	 \\
252	&	100	&	4.97	&	5.04	&	0.235	&	0.04	&	$0^+$	&	$0^+$	&	0	 \\
252	&	101	&	2.93	&	-	&	0.235	&	0.049	&	$$	&	$0^+\#$	&	0	\\
252	&	102	&	0.59	&	0.74	&	0.235	&	0.049	&	$0^+$	&	$0^+$	&	0	 \\
252	&	103	&	-0.68	&	-	&	0.235	&	0.049	&	$$	&	$$	&	0	\\
\hline
\hline
\end{tabular}
\end{table}

\begin{table}
%\caption {Something}
%\label{table:1}
\renewcommand{\tabcolsep}{0.2 pc} % enlarge column spacing
\renewcommand{\arraystretch}{1.2} % enlarge line spacing
\small
\begin{tabular}{ccccccccc}
\hline
\hline
$A_p$ & $Z_p$ & ${\cal T}^{\mathrm{theor}}$ & ${\cal T}^{\mathrm{exp}}$ & $\beta_2$ & $\beta_4$ & $j_\mathrm{p}^{\pi}$ & $j_\mathrm{d}^{\pi}$& $l_\mathrm{min}$ \\
\hline
258	&	106	&	-1.15	&	-	&	0.237	&	0.008	&	$0^+$	&	$0^+$	&	0	 \\
259	&	100	&	8.93	&	-	&	0.227	&	-0.004	&	$1.5^+\#$	&	$(3.5^+)$	&	 2	\\
259	&	101	&	7.91	&	-	&	0.226	&	0.006	&	$3.5^-\#$	&	$(3.5^+)$	&	 1	\\
259	&	102	&	3.69	&	-	&	0.237	&	-0.001	&	$4.5^+\#$	&	$3.5^+$	&	2	 \\
259	&	103	&	3.19	&	-	&	0.237	&	0	&	$4.5^+\#$	&	$(3.5^-)$	&	1	 \\
259	&	104	&	0.46	&	-	&	0.237	&	0	&	$3.5^+\#$	&	$(0.5^+)$	&	4	 \\
259	&	105	&	-1.04	&	-	&	0.246	&	0.002	&	$$	&	$3.5^-\#$	&	0	 \\
\hline
\hline
\end{tabular}
\end{table}

\begin{table}
%\caption {Something}
%\label{table:1}
\renewcommand{\tabcolsep}{0.2 pc} % enlarge column spacing
\renewcommand{\arraystretch}{1.2} % enlarge line spacing
\small
\begin{tabular}{ccccccccc}
\hline
\hline
$A_p$ & $Z_p$ & ${\cal T}^{\mathrm{theor}}$ & ${\cal T}^{\mathrm{exp}}$ & $\beta_2$ & $\beta_4$ & $j_\mathrm{p}^{\pi}$ & $j_\mathrm{d}^{\pi}$& $l_\mathrm{min}$ \\
\hline
266	&	104	&	5.42	&	-	&	0.219	&	-0.039	&	$0^+$	&	$0^+$	&	0	 \\
266	&	105	&	3.72	&	-	&	0.219	&	-0.038	&	$$	&	$$	&	0	\\
266	&	106	&	1.22	&	-	&	0.229	&	-0.037	&	$0^+$	&	$0^+$	&	0	 \\
266	&	107	&	0.10	&	-	&	0.229	&	-0.036	&	$$	&	$$	&	0	\\
266	&	108	&	-2.22	&	-	&	0.229	&	-0.036	&	$0^+$	&	$0^+$	&	0	 \\
266	&	109	&	-2.97	&	-	&	0.239	&	-0.034	&	$$	&	$$	&	0	\\
267	&	104	&	4.87	&	-	&	0.22	&	-0.046	&	$$	&	$$	&	0	\\
\hline
\hline
\end{tabular}
\end{table}

\begin{table}
%\caption {Something}
%\label{table:1}
\renewcommand{\tabcolsep}{0.2 pc} % enlarge column spacing
\renewcommand{\arraystretch}{1.2} % enlarge line spacing
\small
\begin{tabular}{ccccccccc}
\hline
\hline
$A_p$ & $Z_p$ & ${\cal T}^{\mathrm{theor}}$ & ${\cal T}^{\mathrm{exp}}$ & $\beta_2$ & $\beta_4$ & $j_\mathrm{p}^{\pi}$ & $j_\mathrm{d}^{\pi}$& $l_\mathrm{min}$ \\
\hline
275	&	110	&	-2.92	&	-	&	0.222	&	-0.088	&	$$	&	$$	&	0	\\
275	&	111	&	-3.92	&	-	&	0.231	&	-0.087	&	$$	&	$$	&	0	\\
276	&	108	&	2.20	&	-	&	0.201	&	-0.074	&	$0^+$	&	$0^+$	&	0	 \\
276	&	109	&	-0.03	&	-	&	0.221	&	-0.089	&	$$	&	$$	&	0	\\
276	&	110	&	-2.35	&	-	&	0.222	&	-0.089	&	$0^+$	&	$0^+$	&	0	 \\
276	&	111	&	-3.23	&	-	&	0.222	&	-0.088	&	$$	&	$$	&	0	\\
277	&	108	&	4.22	&	-	&	0.183	&	-0.069	&	$1.5^+\#$	&	$$	&	0	 \\
\hline
\hline
\end{tabular}
\end{table}

%\newpage

%\onecolumn
\clearpage

\begin{table}[h]
\caption {RMS errors of the decimal logarithm of $\alpha$-decay half-lives $\delta$ obtained for different models for 344 (tot), 136 even-even(e-e), 84 even-odd (e-o), 76 odd-even (o-e) and 48 odd-odd (o-o) $\alpha$-emitters using our dataset for the ground-state-to-ground-state $\alpha$-transition half-lives.}
\renewcommand{\arraystretch}{1.2} % enlarge line spacing
\vspace{0.2cm}
\begin{tabular}{cccccc}
\hline
\hline
tot& e-e & e-o & o-e & o-o& \\
\hline
0.6248&0.3088&0.7816&0.7621&0.7546& {\it UMADAC} \\
1.0185&0.5165&1.1611&1.3348&1.2568&\cite{dasgupta}\\
1.1130&0.3837&1.4762&1.3688&1.3340&\cite{royer}\\
1.1285&0.3712&1.5425&1.3541&1.3307&\cite{brasil2}\\
1.3813&1.2928&1.4300&1.5607&1.2751&\cite{mnms2}\\
\hline
\hline\\
\end{tabular}
\end{table}

\vspace{-.5cm}
\begin{table}[r]
\caption {RMS errors of the decimal logarithm of $\alpha$-decay half-lives $\delta$ obtained for different models  for $\alpha$-emitters heavier than $^{208}_{82}$Pb using our dataset for the ground-state-to-ground-state $\alpha$-transition half-lives. This dataset contains 144 (tot), 59 even-even(e-e), 33 even-odd (e-o), 34 odd-even (o-e) and 18 odd-odd (o-o) $\alpha$-emitters.}
\renewcommand{\arraystretch}{1.2} % enlarge line spacing
\begin{tabular}{cccccc}
\hline
\hline
tot& e-e & e-o & o-e & o-o &\\
\hline
0.7170&0.3135&0.9521&0.9184&0.8032&{\it UMADAC}\\
1.2326&0.2854&1.8008&1.4748&1.4753&\cite{mnms2}\\
1.2516&0.3861&1.6558&1.5062&1.7615&\cite{dasgupta}\\
1.3410&0.3067&2.0223&1.6186&1.4219&\cite{sp}\\
1.4399&0.2202&2.1371&1.6545&1.8339&\cite{royer}\\
1.4933&0.3701&2.2528&1.6663&1.8292&\cite{brasil2}\\
1.6926&0.2187&2.5050&1.9202&2.2285&\cite{gupta}\\
8.6375&7.6761&9.7315&9.0791&9.3946&\cite{fuji}\\
\hline
\hline\\
\end{tabular}
\end{table}

%\newpage
\clearpage
\begin{figure*}
\includegraphics[width=17.5cm]{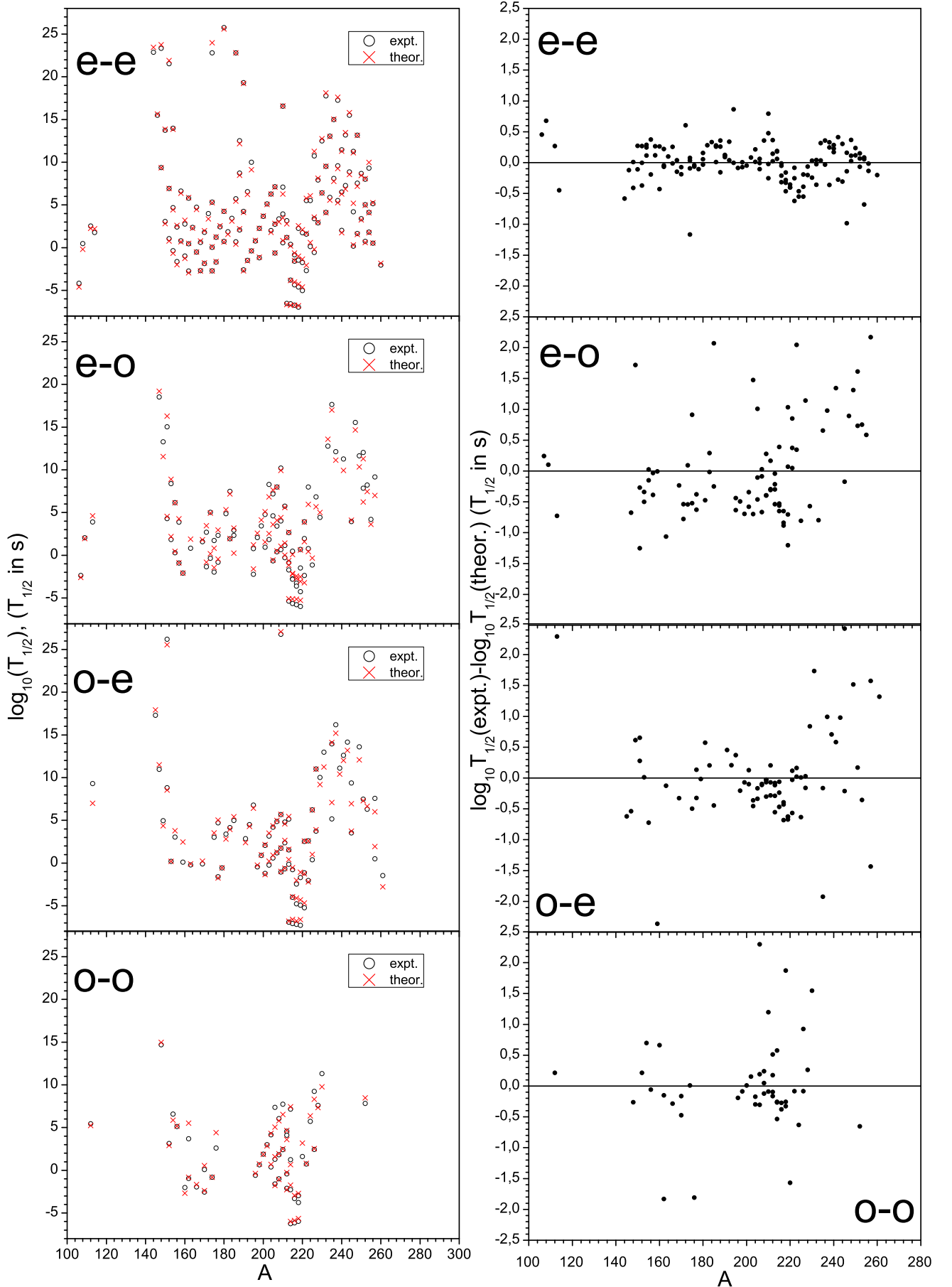}
\caption{(Color online) The difference between the experimental and theoretical values of $\log_{10}(T_{1/2})$ for $\alpha$-decays in even-even (e-e), even-odd (e-o), odd-even (o-e) and odd-odd (o-o) parent nuclei. $T_{1/2}$ values are in s.}
\end{figure*}

%\clearpage
\begin{figure*}
\includegraphics[width=15cm]{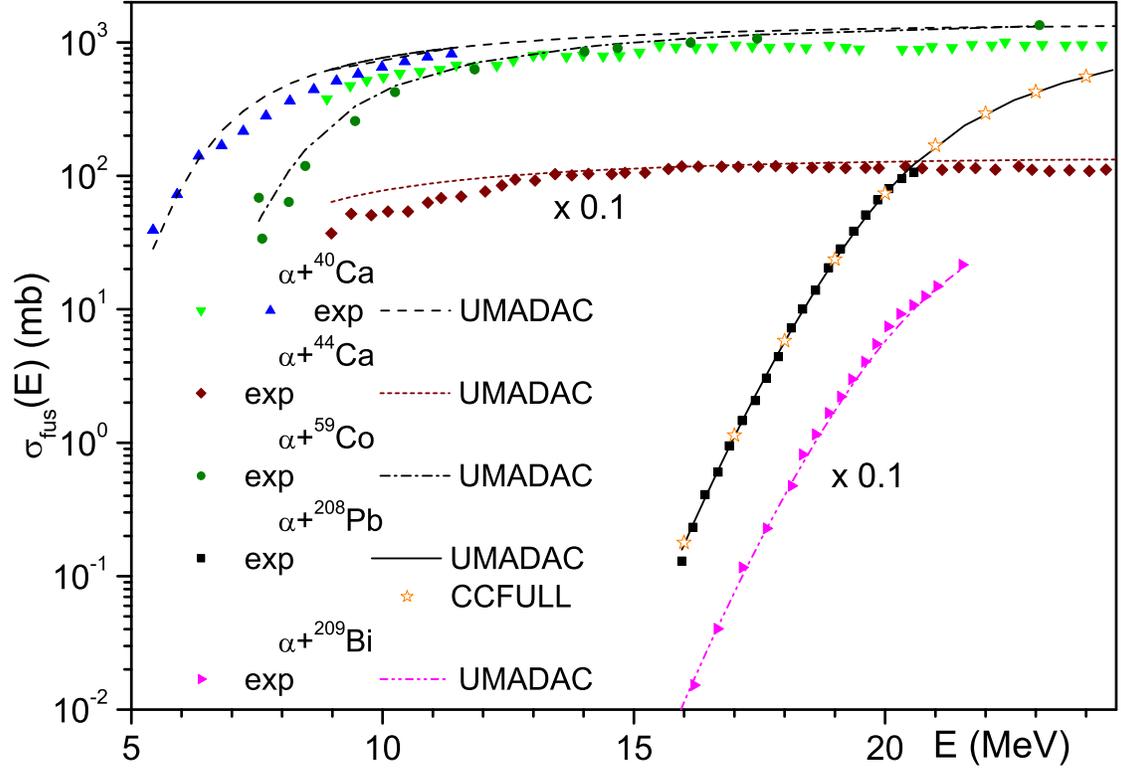}
\caption{(Color online)The experimental and theoretical values of $\alpha$-capture cross-section of $^{40}$Ca, $^{44}$Ca, $^{59}$Co, $^{208}$Pb and $^{209}$Bi.}
\end{figure*}

\end{document}